\documentclass[10pt,journal,compsoc]{IEEEtran}
\ifCLASSOPTIONcompsoc
  \usepackage[nocompress]{cite}
\else
  \usepackage{cite}
\fi
\usepackage[pdftex]{graphicx}
\DeclareGraphicsExtensions{.eps}

\usepackage{authblk}
\usepackage[utf8]{inputenc}
\usepackage{multirow}
\usepackage{balance}
\usepackage{booktabs}
\usepackage{mdframed}
\usepackage{xspace}
\usepackage{listings}
\usepackage[dvipsnames]{xcolor}
\usepackage{numprint}

\definecolor{trycatch}{RGB}{120, 5, 100}
\definecolor{annot}{RGB}{5, 5, 120}
\definecolor{comments}{RGB}{35, 100, 5}
\AtBeginDocument{%
}
\newcommand\chaosmachine{{\sc ChaosMachine}\xspace}
\newcommand\chaosmachinebf{{\sc \textbf{ChaosMachine}}\xspace}

\begin{document}
\title{\huge{A Chaos Engineering System for Live Analysis and Falsification of Exception-handling in the JVM}}
\author[1]{Long Zhang}
\author[2]{Brice Morin}
\author[1]{Philipp Haller}
\author[1]{Benoit Baudry}
\author[1]{Martin Monperrus}
\affil[1]{KTH Royal Institute of Technology, Sweden}
\affil[2]{SINTEF, Norway}
\date{} 
\setcounter{Maxaffil}{0}
\renewcommand\Affilfont{\itshape\small}

\IEEEtitleabstractindextext{
\begin{abstract}
Software systems contain resilience code to handle those failures and unexpected events happening in production.
It is essential for developers to understand and assess the resilience of their systems. Chaos engineering is a technology that aims at assessing resilience and uncovering weaknesses by actively injecting perturbations in production.
In this paper, we propose a novel design and implementation of a chaos engineering system in Java called \chaosmachine. It provides a unique and actionable analysis on exception-handling capabilities in production, at the level of try-catch blocks.
To evaluate our approach, we have deployed \chaosmachine on top of 3 large-scale and well-known Java applications totaling $630k$ lines of code. Our results show that \chaosmachine reveals both strengths and weaknesses of the resilience code of a software system at the level of exception handling.
\end{abstract}

}

\maketitle

\IEEEraisesectionheading{\section{Introduction}\label{sec:introduction}}
\IEEEPARstart{C}{haos} engineering is a new scientific method within software engineering that consists
in specifying and evaluating resilience hypotheses by 1) injecting faults in a production system, 2) observing the impact of such faults, and 3) building new knowledge about the strengths and weaknesses of the resilience of the system~\cite{Basiri:Chaos_Engineering:IEEESoftware}.
The core idea of chaos engineering is active probing:
the chaos engineering system actively injects a controlled perturbation into the production system and observes the impact of the perturbation as well as the reaction of the system under study~\cite{Allspaw:Fault_Injection_in_Production,Chang:Chaos_Monkey_Increasing_SDN_Reliability_Through_Systematic_Network_Destruction,The_Netflix_Simian_Army}. 
This aids developers in gaining confidence in the resilience of their system, and can help them find weaknesses in error handling and disaster recovery routines~\cite{Basiri:Chaos_Engineering:IEEESoftware,Alvaro:Automating_Failure_Testing_Research_at_Internet_Scale}.

Chaos engineering is fundamentally a production testing method that complements other testing methods such as static analysis or staging environments. The key features of chaos engineering~\cite{Gunawi:Failure_as_a_Service,Leesatapornwongsa:The_Case_for_Drill-Ready_Cloud_Computing} are that 1) it does not require reproducing a realistic production environment in a testing setup, and 2) it is focused on investigating resilience hypotheses (and not on functional correctness).
An example of a chaos engineering system is Netflix ChaosMonkey, which randomly shuts down servers to make sure that the overall system is capable of spawning new ones automatically.

In this paper, we present the design and implementation of a novel chaos engineering system called \chaosmachine. Its core novelty and uniqueness is that it considers error-handling capabilities at the fine-grained level of programming language exceptions. While bad exception-handling is known to be the cause of up to 92\% of critical failures \cite{DingYuan:Simple_Testing_Can_Prevent_Most_Critical_Failures}, it remains to be done to apply the chaos engineering vision to exception-handling. The contribution of this paper is \chaosmachine which is capable of revealing the resilience strengths and weaknesses for every try-catch block executed in production.

\chaosmachine is designed around three components. For each service of the system under study, there is a monitoring sidecar (component \#1) and a perturbation
injector (component \#2) attached to it. The monitoring sidecar is responsible for collecting all information needed for the resilience analysis, and the perturbation injector is able to throw a specific exception at runtime. Component \#3 is the chaos controller, which controls all the perturbation injectors and analyzes the information collected by all monitoring sidecars. Eventually, the chaos controller produces a report that gives developers unique and actionable knowledge about the resilience of their system.

We evaluate \chaosmachine by applying it to 3 large-scale and well-known open-source Java applications in the domains of file sharing, content-management system, and e-commerce. All our experiments are conducted in a production-ready environment with end-user-level workload. 
The results show that \chaosmachine is capable of analyzing the resilience of $339$ try-catch blocks located in $212$ Java classes. \chaosmachine successfully identifies the resilient try-catch blocks ($18/339$) that should remain resilient in subsequent versions (according to our definition of resilience presented in \autoref{sec:components}).
It also identifies the weakest try-catch blocks ($34/339$) which are possible debug nightmares when developers try to understand failures happening in production. 

In summary, our main contributions are the following.

\begin{itemize}

\item The conceptual foundations of chaos engineering in the context of exception-handling in Java: 1) the definition of four categories of try-catch blocks according to their resilience characteristics; 2) a systematic procedure to assess resilience of try-catch blocks.

\item A novel system, called \chaosmachine, that assesses exception-handling capabilities in production, based on bytecode instrumentation of Java code. It provides valuable and actionable feedback to the developers. The system is publicly available for future research (\href{https://github.com/KTH/royal-chaos/tree/master/chaosmachine}{https://github.com/KTH/royal-chaos/tree/master/chaosmachine}).

\item An empirical evaluation of \chaosmachine on 3 real-world Java systems totaling $630k$ lines of code, containing $339$ try-catch blocks executed by the considered production traffic. It shows the effectiveness of \chaosmachine to reveal both strengths and weaknesses of a software system's resilience at the exception-handling level.
\end{itemize}

The rest of the paper is organized as follows: \autoref{sec:backgroundOnChaosEngineering} presents the background, \autoref{sec:theDesignOfChaosMachine} and \autoref{sec:evaluation} describe the design and evaluation of \chaosmachine.
 \autoref{sec:relatedWork}  and \autoref{sec:conclusion} discuss the literature and future work.

\section{Background on Chaos Engineering}\label{sec:backgroundOnChaosEngineering}

In this section, we give some background on chaos engineering~\cite{Basiri:Chaos_Engineering:IEEESoftware} for readers who are not familiar with the concept.

\subsection{Brief Overview}

Chaos engineering is a scientific method to verify resilience hypotheses about software systems~\cite{Basiri:Chaos_Engineering:IEEESoftware}. In this context, a hypothesis means that the system is resilient to a specific type of failure. For example, Netflix hypothesizes that their platform is resilient to a server crash.
To investigate this hypothesis, a `chaos experiment' consists of randomly shutting down a virtual machine and checking that this perturbation has no bad influence on the main business metric of Netflix: the number of served video streams per second.

Those resilience hypotheses express properties about the behavior of a whole software system at scale~\cite{Chaos_Engineering_Book:OReilly}. Consequently, the chaos experiments are meant to be run in an environment that is as complex and as unpredictable as the production one. Since it is very challenging to reproduce the complexity and the scale of production environments in a staging environment, chaos experiments are usually run in production. 
By doing this, chaos engineering produces knowledge about the system that is constantly updated to the latest version of the system and its current production workload.
As chaos engineering means experimenting with the system in production, Schermann et al.\cite{Schermann:A_multi-method_empirical_study_on_continuous_experimentation} consider it as one dimension of continuous experimentation.

Let us now discuss a concrete example.
Consider two micro-services interacting with each other to provide a feature. Those two micro-services have error-handling code to deal with problems in the communication link.
Chaos engineering on this system would mean injecting a perturbation in the communication link in production. If the system continues to provide the expected service under this perturbation, the developers gain confidence in the error-handling code. If any perturbation breaks the system's provided features, it means that the developers need to fix the error-handling code. This is the meaning behind the primary idea of chaos engineering: ``\emph{experimenting on a distributed system in order to build confidence in the system’s capability to withstand unexpected conditions in production}''~\cite{Principles_Of_Chaos_Engineering}.

\subsection{Core Concepts}\label{sec:coreConcepts}

Chaos engineering is founded on the following concepts.

A \emph{perturbation} is a change in the application execution flow, or state, or environment; it is made in a pro-active and controlled manner. An example of a perturbation in the system environment is when one cuts down the memory available to the system to see how the application reacts. Perturbations are controlled by a perturbation controller, which may be centralized or distributed depending on the application.

A \emph{hypothesis} is a stipulated relation between a perturbation and observed, or monitored, behaviors.
In the case of a video streaming service, one monitored behavior can be the number of streams started per second. The behaviors of interest can be caught using a wide range of tools:
core business metrics (e.g., number of streams),
execution metrics (e.g., number of executions of a specific method), 
environment metrics like I/O usage, etc.
For example, a hypothesis may be:
in our web page rendering system, if one stops the cache 
(a perturbation simulating that the cache subsystem is broken), the correct content is still delivered to users (monitored behavior).

An \emph{experiment} is the process of validating or falsifying a hypothesis. An experiment includes injecting perturbations into the system, monitoring how the system reacts and inferring validation and falsification. In the example above, an experiment for this hypothesis is:
1) to inject an exception into the page rendering service, and
2) to monitor the system's reaction. 
If there is a difference between the behavior under injection and the normal behavior, the experiment is considered to have \emph{falsified} the hypothesis, where ``difference'' is defined as a distance metric between two observation metrics (examples of distance metrics are given in \autoref{sec:monitoringSidecars}).

\subsection{Basic Chaos Methodology}\label{sec:methodology}
Per~\cite{Principles_Of_Chaos_Engineering}, there are four main steps to apply chaos engineering to a system.

The first three steps are related to designing the hypothesis.
First of all, one must find metrics which capture the essential performance and correctness characteristics of the system's steady state. The steady state is characterized by a range of metric values, with departure from that range meaning the system should be considered impacted. 
Secondly, one defines perturbations which simulate real-world possible events, such as connection timeout, hard drive exhaustion, thread death, etc.

Then one defines two phases: a control phase and an experimental phase. A control phase is a monitoring period without perturbation, while the experimental phase is a study of the system behaviors under perturbation. At the end of these first three steps, the hypothesis and the experiment design are set.

Fourth, one performs the actual experiment, consisting of injecting perturbations into the system and monitoring the metrics. A report is eventually generated by analyzing the differences in the effect of the perturbation between the control phase and the experimental phase. There are two possible outcomes:
1) when a hypothesis is validated, the confidence in the system resilience capability is improved;
2) when a hypothesis is falsified, the issue is reported to the development team, which then has to fix the error-handling code.

Arguably, the most famous chaos engineering system to date is Netflix' ChaosMonkey.
ChaosMonkey is used to randomly shutdown systems that are part of a fleet providing a service in production and to then analyze impact on that system. Developers at Netflix use the number of video starts per second as a metric to define the system's steady state~\cite{Chaos_Engineering_Book:OReilly}. In this context, 
1) an hypothesis is that one instance terminating abnormally has no influence on the number of served videos; 
2) a perturbation is ChaosMonkey shutting down a specific instance; 
3) a chaos experiment is the whole procedure of applying ChaosMonkey and analyzing the system's behavior to validate or falsify the hypothesis.

\subsection{Differences Between Traditional Fault Injection and Chaos Engineering}\label{sec:differences}

Chaos engineering \cite{Chaos_Engineering_Book:OReilly} and fault injection \cite{natella2016assessing} are closely related, as they both seek to perturb the runtime execution of a system under study. We now discuss key differences between both, as well as the main benefits and risks related to chaos engineering.

\textbf{Differences} The major difference is that chaos engineering is fault injection \emph{in production}, while traditional fault injection is usually done in a testing, off-line environment. This has major implications. First, fault injection in production imposes constraints on the overhead of fault injection: production systems can be slowed down only to a reasonable extent.
Second, it means that the reasoning on the results of fault injection in production has to be done with the oracles and monitoring information that are available in production. On the contrary, traditional fault injection may have large overhead or may assume some level of observability that is not available in production for security or privacy reasons. For example, observing the content of CPU registers is typically not available on a production system. 

\textbf{Benefits} The benefits of doing fault injection \emph{in production} are:
1) one has access to the system at scale with its full complexity and size;
2) one has access to the production inputs and workloads.
On the contrary, a testing environment for fault injection is usually smaller than the production system, which may hide some defects or the simulated workloads may be restricted or artificial. This results in a lower external validity.  In some cases, it may be impossible to set up a testing environment for large distributed systems~\cite{Alvaro:Abstracting_the_Geniuses_Away_from_Failure_Testing}. 

\textbf{Risk} Doing fault injection \emph{in production} is risky \cite{Chaos_Engineering_Book:OReilly}: it may result in data corruption, degraded user experience, and financial losses. Consequently, extra care is put in the engineering of mitigating the impact of the perturbations. Note the high interest of industry for chaos engineering shows that the trade-off benefits/risks are interesting in certain well-defined business cases.

\section{Design of Chaos System for Exception-Handling}\label{sec:theDesignOfChaosMachine}

This section presents our system for controlled chaos engineering in the Java Virtual Machine, called \chaosmachine. Its core novelty is that it does chaos engineering at the level of exception handling and try-catch blocks, which is more fine-grained than all chaos engineering systems we are aware of.

\subsection{Overview}\label{sec:overview}

The goals of \chaosmachine are:
\begin{enumerate}
    \item allow developers to specify hypotheses directly in their source code,
    \item falsify hypotheses, and
    \item discover hypotheses.
\end{enumerate}

Goal 2 is the classical goal of chaos engineering systems~\cite{Basiri:Chaos_Engineering:IEEESoftware}, and the other two goals are the key contribution of this paper.

\textbf{Hypotheses.}
\chaosmachine considers error-handling hypotheses in Java applications.
We focus on try-catch blocks and consider the exception type that is caught (both checked and unchecked).\footnote{If a checked exception is only passed (``throws'' declaration in Java), there is no resilience involved and remains out of our scope.}

We define the following four chaos engineering hypotheses at the level of try-catch blocks, from the most beneficial to the most problematic:
\begin{itemize}
\item \emph{Resilience hypothesis}. A try-catch block is said to be resilient if the observable behavior of the catch block, executed upon exception, is equivalent to the observable behavior of the try-block when no exception happens~\cite{Cornu:Exception_Handling_Analysis_and_Transformation_Using_Fault_Injection}. \autoref{lst:resilience_hypothesis} presents an example of a try-catch block which meets the resilience hypothesis.
\item \emph{Observability hypothesis}. A try-catch block is said to be observable if an exception caught in the catch block results in user-visible effects, see \autoref{lst:observability_hypothesis}.
\item \emph{Debug hypothesis}. A try-catch block is said to be debuggable if an exception caught in the catch block results in an explicit message in the application logs, see \autoref{lst:debug_hypothesis}.
\item \emph{Silence hypothesis}. A try-catch block is said to be silent if it fails to provide the expected behavior upon exception while providing no troubleshooting information whatsoever, i.e., it is neither observable nor debuggable. If the silent try-catch block later causes a user-visible failure, it would be extremely hard for the developers to understand that the root cause is the silent try-catch block, and to fix the failure accordingly, see \autoref{lst:silence_hypothesis}.
\end{itemize}

\begin{lstlisting}[caption=Try-catch Satisfying the Resilience Hypothesis,label=lst:resilience_hypothesis,belowskip=-1em]
state = SystemState.A;
try {
    ... // an error is thrown
    state = SystemState.B;
} catch (
    @ChaosMachine(hypoth=Hypoth.RESILIENT) 
    Exception e
  ) {
    ... // handles the exception
    state = SystemState.B;
}
// After leaving the try-catch, the state stays the same
\end{lstlisting}

\begin{lstlisting}[caption=Try-catch Satisfying the  Observability Hypothesis,label=lst:observability_hypothesis,belowskip=-1em]
try {
    contentsToUsers.add("content A");
    contentsToUsers.add("content B");
} catch (
    @ChaosMachine(hypoth=Hypoth.OBSERVABLE) 
    Exception e
  ) {
    contentsToUsers.add("content C");
}
render(contentsToUsers);
// When exception occurs, contents for users are different
\end{lstlisting}

\begin{lstlisting}[caption=Try-catch Satisfying the Debug Hypothesis,label=lst:debug_hypothesis,belowskip=-1em]
try {
    ...
} catch (
    @ChaosMachine(hypoth=Hypoth.DEBUG) 
    Exception e
  ) {
    ...
    // Log troubleshooting information
    Logger.error("...domain specific information...");
}
\end{lstlisting}

\begin{lstlisting}[caption=Try-catch Satisfying the  Silence Hypothesis,label=lst:silence_hypothesis]
state = SystemState.A;
try {
    state = SystemState.B;
    contentsToUsers.add("content A");
} catch (
    @ChaosMachine(hypoth=Hypoth.SILENT) 
    Exception e
  ) {
    state = SystemState.C;
    contentsToUsers.add("content A");
    // Nothing about the exception is logged
}
render(contentsToUsers);
// Users are not aware of the error, but system state is different when an exception occurs, which may lead to other exceptions.
\end{lstlisting}

\textbf{Experiments.} \chaosmachine performs two kinds of experiments, falsification experiments and exploration experiments that we now explain.

\emph{Falsification experiments} aim at validating or falsifying a hypothesis about the behavior of a try-catch block. This hypothesis can be stated upfront by developers or can be discovered through exploration experiments. When they are stated by developers, it is in the form of an annotation \texttt{@ChaosMachine} defined directly on the exception within a catch statement, as illustrated in the previous scripts. This allows developers to explicitly emit hypotheses on specific try-catch blocks, which they deem of critical importance, based on their knowledge. Those annotations are processed to produce a textual configuration file listing the try-catch block identifiers and their hypothesized status per exception (one hypothesis per line).
The internal format used by \chaosmachine is as follows: \texttt{Foo.java:42:NullPointerException "observable"} where the identifier is the tuple composed of the name of the class, the line number of the beginning of the try block, and the caught exception followed by space and the hypothesized status.\footnote{It may happen that there are multiple try-blocks on the same line, if this is the case, \chaosmachine supports more advanced unique identifiers.} 
The output lists the results: \texttt{Foo.java:42:NullPointerException "observable": FALSIFIED}. Just like hypotheses discovered by \chaosmachine, those developer-specific hypotheses will be either verified or falsified, confirming or invalidating the knowledge developers have on those annotated try-catch blocks. The main benefit of annotating try-catch blocks upfront is that \chaosmachine will be able to precisely relate back the specific lines in the source code where an hypotheses were falsified.

\emph{Exploration experiments}. They aim at monitoring the behavior of try-catch blocks under perturbation in order to discover new hypotheses. After an exploration experiment, \chaosmachine outputs a file with the try-catch block identifiers and their status, for example, \texttt{Bar.java:42:NullPointerException "silent"}.

\textbf{Modes.} When \chaosmachine performs exploration experiments, it is said to be in \emph{exploration mode}. When \chaosmachine performs falsification experiments, it is in \emph{falsification mode}. Finally, when \chaosmachine does not introduce chaos, it is simply in \emph{observation mode}.

\subsection{Input to \chaosmachinebf}
\chaosmachine works on arbitrary software written in Java, no manual change is required in the code.
To use \chaosmachine, the application is deployed in production as usual, \chaosmachine is attached to it in an automated manner, in observation mode by default.
Optionally, developers can also feed \chaosmachine with manually-written hypotheses.
The controller of \chaosmachine, described in \autoref{sec:chaosController}, determines what must be done for perturbation and monitoring according to the current mode.

\subsection{Architecture of \chaosmachinebf}\label{sec:components}
\begin{figure*}
\centering
\includegraphics[width=17.5cm]{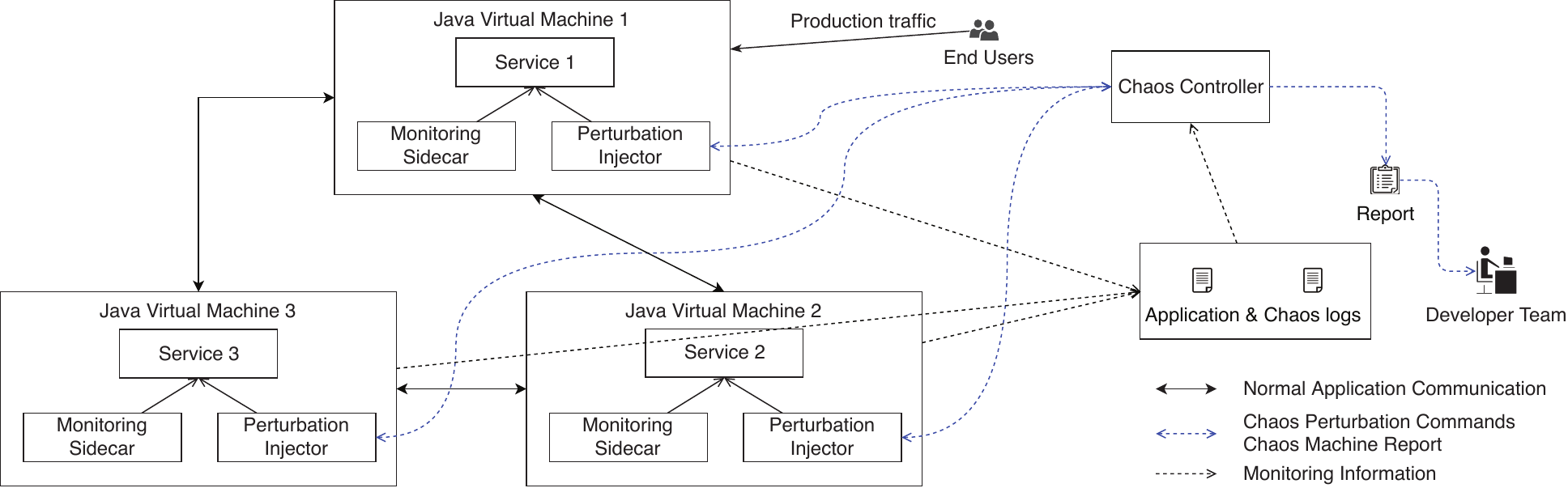}
\caption{The components of \chaosmachine}\label{fig:components}
\end{figure*}

\begin{table*}
\centering
\caption{Interplay between the 3 components and the 3 modes of \chaosmachine}\label{tab:usagesInModes}
\begin{tabular}{l p{4.5cm}p{4.5cm}p{4.5cm}}
\toprule
& {\bf Observation Mode} & {\bf Exploration Mode} & {\bf Falsification Mode}\\
\midrule
{\it Monitoring Sidecar} & Monitors all the relevant execution information& Monitors how the system reacts according to a perturbation& Monitors whether an hypothesis is falsified\\
\hline
{\it Perturbation Injector} & Not active& Injects a specific perturbation& Injects a specific perturbation\\
\hline
{\it Chaos Controller} & Deactivate all the perturbation injectors to keep the system running as usual& Controls perturbation injectors to conduct a sequence of chaos experiments so as to discover new hypotheses& Controls perturbation injectors according to a specific hypothesis\\
\bottomrule
\end{tabular}
\end{table*}

\autoref{fig:components} presents the main components of \chaosmachine and their interactions. \chaosmachine is meant to be deployed on any modern Internet application, such as search engines or transaction systems. Those applications typically are distributed over several different servers, where the servers either provide different services (as shown in the figure with three different services), or provide redundancy and elasticity for the same service.
Per the best practices, and without loss of generality, all services are considered deployed in separate virtual machines for the sake of isolation.

\chaosmachine attaches a monitoring sidecar and a perturbation injector into each service. 
The monitoring sidecar (\autoref{sec:monitoringSidecars}) collects information to study the outcome of chaos experiments. 
The perturbation injector (\autoref{sec:perturbationInjectors}) is responsible for injecting perturbations according to a given perturbation model. A perturbation model is a set of rules that describe when, where and what perturbations are to be injected. For example, a perturbation model may be: inject an \texttt{IOException} at the first line of method \texttt{foo()}, when this method is reached for the first time.\footnote{This is related to a ``fault model'' in the fault injection literature \cite{natella2016assessing}. ``Fault model'' is a more general concept that is also used in electrical engineering, while perturbation model is specifically used by chaos engineering.}
The chaos controller (\autoref{sec:chaosController}) is a standalone component that is separated from the application services, and it has three responsibilities: 
1) controlling the behavior of perturbation injectors;
2) aggregating monitoring information from each monitoring sidecar;
3) generating a report for the developers about the quality of error-handling in their code, which contains novel and actionable feedback about error-handling in production. We further describe these outputs in~\autoref{sec:outputForDevelopers}.

\subsubsection{Monitoring Sidecars}\label{sec:monitoringSidecars}
Chaos engineering consists of studying the influence of perturbations on the system behavior, as captured by metrics~\cite{Principles_Of_Chaos_Engineering}. 
The main role of the monitoring sidecar is to collect behavioral metrics at runtime.

\textbf{Metrics}
Metrics are used by \chaosmachine to capture a behavioral difference under perturbation.
There are predefined metrics available by default in the system: 
\begin{itemize}
\item The proportion of each process exit status to track whether a service has exited normally or not (crash).
\item The proportion of each HTTP status for web requests.
\item A set of standard operating system metrics including CPU usage, memory usage, and peak thread number.
\item The frequency of each unique log line in the application logs.
\end{itemize}
If the user wants to add its own metrics, this is possible through ``user-defined metrics'' that are domain-specific. She is responsible to push the metric data to an endpoint on the chaos controller (for instance, a video streaming service may push the number of streams started per second). 

\textbf{Distance Measures} The collected metrics are used to determine if a behavior under perturbation is acceptable or not. All default metrics come with a standard acceptance criteria, eg $n_{exit != 0} > 0$ for process exit statuses.

\textbf{Propagation and Observability} Note that the injection points and the observation points are uncorrelated, meaning that the point of observation may be far away from the catch block (both in space in the program and in time in the execution). In other terms, the monitoring sidecar checks whether the exception propagates in the program.

\textbf{Injection Monitoring}
The monitoring sidecar finds the try-catch blocks dynamically, as they are loaded into the JVM. For each try-catch block the monitoring sidecar notes:
1) the position in the code,
2) the type of the caught exception, 
3) the number of executions in observation mode
4) the number of executions in exploration mode.
5) the current call stack for each injection. 

The features of the monitoring sidecars are important to lower the risk of conducting chaos experiments: the more monitoring information \chaosmachine collects, the more accurate it is to detect severe impacts of a specific perturbation injection.

\subsubsection{Perturbation Injectors}\label{sec:perturbationInjectors}

The main responsibility of a perturbation injector is to generate a specific perturbation when the chaos controller sends the corresponding command, i.e., throwing an exception at a specific line.
The injector is added using automated code instrumentation. In order to mitigate the influence of chaos experiments, each injector can be activated and deactivated individually.

\autoref{lst:shortScircuitExample} gives an example of this perturbation injector. There are two try-blocks in this code snippet, Exception1 and Exception2 might happen in the first try-block during the execution of the omitted code at line 6, and Exception3 might happen in the second try-block at line 12. Consequently, there are three injection points in total, corresponding to each caught exception type. For each perturbation, a unique key is generated based on the location of a try-block and the type of exception it catches. The beginning of this try-block is then instrumented with a call to \texttt{ChaosMachine.throwOrNot}, using this key as an argument. In this way, every injector can be controlled separately. For the sake of clarity, \autoref{lst:shortScircuitExample} shows the effect of the instrumentation on the source code. In practice, however, the developer does not see the instrumented code, since it happens dynamically when classes are loaded, through bytecode manipulation. This means that there is no need for the developer to declare variables or update the list of imported packages. When an injector is activated in exploration or falsification mode, it throws the corresponding exception.

\begin{lstlisting}[float,caption={The Application Code is Automatically Transformed for Injecting Perturbations (Line 2-5, 10-11 Are Injected by \chaosmachine)},label=lst:shortScircuitExample,belowskip=-1em]
try {
    // injector #1, type: Exception1
    ChaosMachine.throwOrNot(key1);
    // injector #2, type: Exception2
    ChaosMachine.throwOrNot(key2);
    ...original code...
} catch (Exception1 e1) {
    ...original code...
    try {
        // injector #3, type: Exception3
        ChaosMachine.throwOrNot(key3);
        ...original code...
    } catch (Exception3 e3) {
        ...original code...
    }
} catch (Exception2 e2) {
    ...original code...
}
\end{lstlisting}

\subsubsection{Chaos Controller}\label{sec:chaosController}
The chaos controller orchestrates the behavior of all perturbation injectors. To reason about the impact of perturbations, it aggregates the information provided by the monitoring sidecars. According to the monitoring data, the chaos controller discovers new hypotheses or falsifies existing hypotheses on the software system under study.

\paragraph{Hypothesis discovery}
Hypothesis discovery consists of proactively analyzing every executed try-catch block. To do so, the chaos controller iterates over them one after the other, to activate the corresponding perturbation injector, and then analyzes all output as captured by the monitoring sidecars. If the discovered hypothesis is considered as acceptable by the developers, the new hypothesis is saved permanently, so that \chaosmachine could conduct falsification experiments on it later on. Hypothesis discovery is fully automated if the default metrics are used. If the developer wants to limit hypothesis discovery to a specific package, she can configure the controller with a list of packages to be included.

\paragraph{Hypothesis falsification}
To falsify a hypothesis, the chaos controller activates the specific perturbation injector which corresponds to the try-catch block in the hypothesis. For example, it only activates the injector of the try-catch block on line 42 of Foo.java, and keeps it activated for a specific number of executions (e.g. for $n=5$ executions of the try block). Then, the controller analyzes the information recorded by the monitoring sidecars and reports whether the injected perturbation has broken the hypothesis under consideration.

In both cases, the chaos controller is configured by a configurable time window: it activates a single perturbation point for a certain duration (e.g., 1 second). In order to minimize the impact of a perturbation experiment, the chaos controller never activates more than one exception injector at a time.

\subsection{Output for Developers}
\label{sec:outputForDevelopers}

\chaosmachine produces a report for developers, containing the hypotheses validated or falsified for each try-catch block, sorted according to their criticality. This provides developers with an overview of the resilience of their system. Silent catch blocks are usually the ones that require the most urgent attention, as they hurt the resilience and/or debuggability of the system. Resilient catch blocks help the resilience, by keeping the system running even when certain exceptions happen.

\subsection{Implementation}
\chaosmachine is written in Java in $2.1k$ lines of code. Both the monitoring sidecars and the perturbation injectors are woven into the application services using a JVM agent~\cite{Ghosh:Bytecode_Fault_Injection_for_Java_Software}.
The agent adds the monitoring and injection code using binary code transformation with the ASM library.\footnote{\url{http://asm.ow2.org}} The code instrumentation happens when a class is loaded into the JVM. With the help of the ASM library, it is possible to iterate over all methods of the loaded class and locate every try-catch block. Then, a piece of byte-code is inserted at the beginning of each try-catch block; it communicates with the chaos controller and throws a corresponding exception. We leverage the Java Management Extension (JMX) mechanism\footnote{\url{https://en.wikipedia.org/wiki/Java_Management_Extensions}} for monitoring sidecars, to retrieve observation metrics like CPU utilization and memory usage.
The chaos controller is a standalone service communicating with the monitoring sidecars and the injectors using sockets.
For the sake of open science, the code is made publicly available at \href{https://github.com/KTH/royal-chaos/tree/master/chaosmachine}{https://github.com/KTH/royal-chaos/tree/master/chaosmachine}.

\section{Evaluation}\label{sec:evaluation}
In our evaluation, we apply \chaosmachine to 3 different real-world Java projects, including TTorrent (a peer-to-peer file downloading tool based on the BitTorrent protocol), Broadleaf (a web-based commercial system) and XWiki (a web-based wiki system). Following the chaos engineering principles, all applications are set up in a production environment. In the following, we present the protocol, experimental results, case studies, and discussions for each project.

All the experiments rely on the same basic principles: 1) the chaos controller evaluates try-catch blocks one by one in the loading order of classes, 2) it only activates one perturbation injector at a time and 3) the activation duration for every injector is identical for all tasks. Once the application is set up, \chaosmachine verifies and evaluates try-catch blocks following the procedure described in~\autoref{sec:overview}.

\subsection{Evaluation on TTorrent}

\subsubsection{Overview of the BitTorrent protocol}
BitTorrent is a peer-to-peer data transfer protocol, which is widely used to download files over the Internet. The core concept of the BitTorrent protocol is that users who want to download a file also serve other users the file parts that they have already downloaded. There are 4 parts in a typical file transfer scenario with the BitTorrent protocol: 1) a torrent file which includes information about the shared files and the tracker servers; 2) several tracker servers which receive client registrations and announce resource information to new clients; 3) clients who want to download files, and then get the torrent files and register their download status with the tracker server;  4) clients who have already downloaded files and provide pieces of the files to others (called the ``seeders'').

\subsubsection{Experiment protocol}\label{sec:experimentProtocolTTorrent}
In this experiment, we consider the Bittorrent client called TTorrent (version \href{https://github.com/mpetazzoni/ttorrent/tree/ttorrent-1.5}{1.5}), written in Java. It is built as a .jar file and can be used on the command line. We attach \chaosmachine to this client, and then use it to download \href{https://www.ubuntu.com/download/alternative-downloads}{ubuntu-14.04.5-server-i386.iso}, a Linux distribution installer of 623.9MB from the Canonical company. This means that we use tracker servers from somewhere else on the Internet, and use many seeders that are providing pieces of the downloaded file.

First, we classify try-catch blocks in TTorrent by refining the four hypotheses discussed in~\autoref{sec:overview} with the combining monitoring metrics specific to this application domain.

\begin{itemize}
\item \emph{Resilient try-catch block}. Despite injected exceptions in this block, the client successfully downloads the file and exits normally. The chaos controller also detects some error messages in the application log. Even though there is an exception thrown in the try-catch block, the application still fulfills the user's requirement correctly. This kind of try-catch block contributes to the application's resilience, as the application still supports the users' requests even though the entire logic of the try-block has been discarded.
\item \emph{Observable try-catch block}. A try-catch block is said observable if the client directly crashes or exits with an error message under perturbations, i.e., the perturbation in this try-catch block causes user-visible behaviors of the client.
\item \emph{Debuggable try-catch block}. A try-catch block is said debuggable if the system metrics become abnormal or the exception information is captured in application logs when an exception is injected. The information is useful for developers to debug and improve the system's resilience.
\item \emph{Silent try-catch block}. When an exception occurs in this block, the client does not download the file and just keeps running indefinitely. Worse still, there is not any error information about the injected error. This is a bad case for both users and developers:
users are not made aware that the download is stalled and
developers have no feedback whatsoever about the problem. Developers can improve them so as to be able to detect and debug such a problem if it happens naturally in production.
\end{itemize}

Then, in an initial observation mode, the client downloads the full file once until successful completion. During this phase, \chaosmachine analyzes the client's behavior. 

Next, for the covered try-catch blocks, \chaosmachine executes the procedure defined in~\autoref{sec:overview} while re-downloading the file, and gathers the data shown in~\autoref{tab:tcInfoInTTorrentClient}.
In exploration mode, the perturbed clients might not be able to exit normally, so \chaosmachine keeps the client alive for at most 300 seconds. After this delay, the client is killed and information is logged indicating that the client was killed after this timeout. 

\subsubsection{Experimental results}\label{sec:results-ttorrent}

\begin{table*}[!tbp]
\centering
\scriptsize
\caption{The Results of Chaos Experimentation With Exception Injection on 27 Try-catch Blocks in the TTorrent Bittorrent Client}\label{tab:tcInfoInTTorrentClient}
\begin{tabular}{l p{1.2cm}rrrrrrrr}
\toprule
Try-catch block information& Execution\newline Obse./Expl.& Log.& Downl.& Exit status& Sys. metrics& RH& OH& DH& SH\\
\midrule
BEValue/getBytes,ClassCastException,0& 41 / 1& yes& no& crashed& -& & x& x& \\
BEValue/getNumber,ClassCastException,0& 15 / 1& yes& no& crashed& -& & x& x& \\
BEValue/getString,ClassCastException,0& 37 / 1& yes& no& crashed& -& & x& x& \\
BEValue/getString,UnsupportedEncodingException,1& 37 / 1& yes& no& crashed& -& & x& x& \\
ClientMain/main,CmdLineParser\$OptionException,0& 1 / 1& yes& no& crashed& -& & x& x& \\
ClientMain/main,Exception,1& 1 / 1& yes& no& crashed& -& & x& x& \\
Announce/run,AnnounceException,0& 1 / 60& yes& no& stalled& -& & x& x& \\
Announce/run,InterruptedException,2& 1 / 760& no& yes& normally& threads+& & & x& \\
Announce/run,InterruptedException,3& 1 / 1& no& yes& normally& no diff& x& & & \\
Announce/run,AnnounceException,4& 1 / 1& yes& yes& normally& no diff& x& & x& \\
Announce/stop,InterruptedException,0& 1 / 1& no& yes& normally& no diff& x& & & \\
ConnectionHandler/run,SocketTimeoutException,0& 1290 / \newline 1030& no& yes& normally& no diff& x& & & \\
ConnectionHandler/run,IOException,1& 1290 / 1& yes& yes& stalled& cpu+& & & x& \\
ConnectionHandler/run,InterruptedException,2& 1290 / 2& yes& no& stalled& no diff& & & x& \\
ConnectionHandler/stop,InterruptedException,0& 1 / 1& no& yes& normally& no diff& x& & & \\
ConnectionHandler\$ConnectorTask/run,Exception,0& 50 / 50& yes& no& stalled& no diff& & & x& \\
Handshake/craft,UnsupportedEncodingException,0& 50 / 48& yes& no& stalled& no diff& & & x& \\
PeerExchange/send,InterruptedException,0& 90763 / \newline 210& no& no& stalled& no diff& & & & x\\
PeerExchange/stop,InterruptedException,0& 46 / 44& no& yes& normally& no diff& x& & & \\
PeerExchange\$OutgoingThread/run,InterruptedException,0& 90805 / \newline 32984841& no& no& stalled& cpu+& & & x& x\\
PeerExchange\$OutgoingThread/run,InterruptedException,1& 90763 / \newline 288& no& no& stalled& no diff& & & & x\\
PeerExchange\$OutgoingThread/run,IOException,2& 90805 / 43& yes& no& stalled& no diff& & & x& \\
PeerExchange\$OutgoingThread/run,IOException,3& 90763 / 46& yes& no& stalled& no diff& & & x& \\
Piece/validate,NoSuchAlgorithmException,0& 2564 / \newline 5427& yes& no& stalled& cpu+& & & x& \\
HTTPAnnounceRespMessage/parse,InvalidBEncodingException,0& 3 / 30& yes& no& stalled& no diff& & & x& \\
HTTPAnnounceRespMessage/parse,InvalidBEncodingException,1& 3 / 30& yes& no& stalled& no diff& & & x& \\
HTTPAnnounceResponseMessage/parse,UnknownHostException,2& 3 / 30& yes& no& stalled& no diff& & & x& \\
\hline
total: 27 / 52& 460626 / \newline 32992950& 18& 8& 7& 4& 6& 7& 20& 3\\
\bottomrule
\end{tabular}
\end{table*}

\autoref{tab:tcInfoInTTorrentClient} reads as follows: there are 27 try-blocks covered by the production traffic, i.e., the code in the try-blocks is executed while the client is downloading the file. Each row contains the information of one try-block. The first column is the basic information about each try-catch block, including the class and method names, caught exception type and a number which is used to identify different catch blocks when there is more than one catch block for a single try-block. The second column records the number of executions, in both observation mode and exploration mode. The third column indicates whether the developers have logged the exception in their application logs when such an exception is caught. The forth column shows whether the client has successfully downloaded the file when exceptions are injected in this try-block. The fifth column records the client's exit status. The sixth column indicates differences in system metrics (if any) between the observation mode and the exploration mode. Finally, the last four columns indicate how this try-catch block meets our pre-defined four hypotheses. Since injected exceptions change the execution flow of the application, the number of executions in analysis mode and exploration mode are not necessarily the same.

Take the first row as an example, it shows that there is a try-catch block in the \texttt{getBytes} method inside the \texttt{BEValue} class, which handles a \texttt{ClassCastException}. Through the entire process of downloading the file, it is executed 41 times. When the perturbation injector throws a  \texttt{ClassCastException} exception at the beginning of the try-block, the client does not download the file and crashes. The chaos controller also detects that a specific error message is logged in the application log before its crash. Based on these behaviors, this try-catch block validates the observability hypothesis (OH) and debug hypothesis (DH).

In total, there are 27 try-catch blocks covered by this file-download operation in production. Some of them are executed only once, others up to 90805 times (cf. Column Execution Anal. of~\autoref{tab:tcInfoInTTorrentClient}). 
This information is very important for the developer.
Thanks to \chaosmachine, the developer is able to identify:
6 resilient try-catch blocks,
7 observable try-catch blocks,
20 debuggable try-catch blocks,
and 3 silent try-catch blocks.

\subsubsection{Case studies}

In the following, we detail 4 case studies.

\begin{lstlisting}[float,caption=ClassCastException in BEValue/getBytes,label=lst:ttorrentCase1, belowskip=-1em]
public byte[] getBytes() throws InvalidBEncodingException {
    try {
        return (byte[])this.value;
    } catch (ClassCastException cce) {
        throw new InvalidBEncodingException(cce.toString());
    }
}
\end{lstlisting}

\autoref{lst:ttorrentCase1} shows a part of the \texttt{getBytes} method, containing a single try-catch statement. This try-catch statement is executed 41 times.
When perturbed with an exception injection, the chaos controller verifies that two core hypotheses are validated in production: the exception is logged, and the client exits with an error status. The developer has no further action to take because this try-catch is both observable and debuggable.

\begin{lstlisting}[float,caption=InterruptedException in Announce/run,label=lst:ttorrentCase2,belowskip=-1em]
while (!this.stop) {
    ...
    try {
        Thread.sleep(this.interval * 1000);
    } catch (InterruptedException ie) {
        // Ignore
    }
}
\end{lstlisting}

\autoref{lst:ttorrentCase2} shows the \texttt{run} method in class \texttt{Announce}. The try-block is a piece of code running in a sub-thread. The announce thread starts by making the initial ``started'' announce request to register on the tracker and get an interval value. In observation mode, the try-catch block is executed once. However, in the exploration mode with exception injection, the try-catch block is executed 760 times. Indeed, due to the skip of the \texttt{Thread.sleep}, the while loop runs more times before reaching its objective.
When the perturbation injector injects the exception, the catch-block simply ``swallows'' this exception and does not do anything to handle the exception.
This results in using more computing resources.
As shown in the comment, the developer knows about this behavior. However, thanks to \chaosmachine, she is made aware that ignoring the exception is not good for performance, and she is even given a quantitative measurement (per the system metrics collected by the monitoring sidecar).

\begin{lstlisting}[float,caption=AnnounceException in Announce/run,label=lst:ttorrentCase3,belowskip=-1em]
if (!this.forceStop) {
    ...
    try {
        this.getCurrentTrackerClient().announce(event, true);
    } catch (AnnounceException ae) {
        logger.warn(ae.getMessage());
    }
}
\end{lstlisting}

\autoref{lst:ttorrentCase3} is also from the \texttt{run} method in the \texttt{Announce} class. The exception type is \texttt{AnnounceException} and this try-catch block is executed once in observation mode, and once in the exploration mode. When the perturbation injector injects the exception, the file is still correctly downloaded. Once the client finishes the download, it exits with a normal exit code, and some error messages about this exception appear in the application log. In this case, the try-catch block successfully blocks AnnounceException to break the system. Even though there is only a logging action in the catch block, our manual analysis has revealed that developers have built the resilience mechanism outside this particular catch block.
Thanks to \chaosmachine, the developer has gained confidence in this specific catch block's exception-handling capability.

\begin{lstlisting}[float,caption=InterruptedException in PeerExchange/send,label=lst:ttorrentCase4,belowskip=-1em]
public void send(PeerMessage message) {
    try {
        this.sendQueue.put(message);
    } catch (InterruptedException ie) {
        // Ignore exception
    }
}
\end{lstlisting}

\autoref{lst:ttorrentCase4} shows method \texttt{send} in class \texttt{PeerExchange}. It is executed \numprint{90763} times in observation mode and 210 times in exploration mode. The method invocation \texttt{this.sendQueue.put(message)} at line 3 acquires a lock unless the current thread is interrupted. Since this method invocation may raise a checked \texttt{InterruptedException}, developers need to wrap it with a try-catch block. Indeed, Java enforces that checked exceptions have to either be declared using a \texttt{throws} keyword in the method signature, or be caught and handled using a try-catch block, as shown. When the perturbation injector injects an \texttt{InterruptedException}, the client just keeps running until some external entity (the user or \chaosmachine) kills the process. No information is logged in the application logs. This means that, when this exception happens naturally, users have absolutely no debug information to give to developers. Here, \chaosmachine helps the developer to identify ``nightmare'' debug cases of the form of purely silent try-catch blocks. Based on the \chaosmachine report, the developer is urged to change the exception-handling behavior.

\subsubsection{Falsification on next version}
It is of utmost importance that the resilience capabilities do not degrade over time.
We try to falsify all hypothesis in a version 
of TTorrent (\href{https://github.com/mpetazzoni/ttorrent/commit/081bab49f7928679217d4fd937456f69b6ab7da2}{1.6}) that is subsequent to the analyzed one, with the same protocol.
The result is that no hypothesis discovered on version 1.5 is falsified on version 1.6, which means that the resilient try-catch blocks are still capable of handling unanticipated exceptions and keeping the system steady.

\begin{mdframed}
Main result of the TTorrent experiment: In a real-world production usage, \chaosmachine identifies $6$ resilient try-catch blocks and $3$ silent ones in the TTorrent client. Each silent try-catch block indicates a potential debug case that would be extremely difficult to fix (no visible behavior, no log can be provided by the user). \chaosmachine precisely detects those silent try-catch blocks and reports them to the developer. 
In subsequent versions, \chaosmachine verifies that the $6$ resilient try-catch blocks remain resilient thanks to falsification experiments. 
\end{mdframed}

\subsection{Evaluation on XWiki}
\subsubsection{Introduction of XWiki}
XWiki is a widely-used open-source wiki system developed in Java, and is active over the past 14 years. XWiki requires external dependencies like a database server and a web application server.

\subsubsection{Experiment protocol}\label{sec:experimentProtocolXWiki}
We use a full-fledged production setup of XWiki version \href{http://download.forge.ow2.org/xwiki/xwiki-9.11.1.war}{9.11.1}, which is deployed into \href{https://archive.apache.org/dist/tomcat/tomcat-8/v8.5.29/}{Tomcat-8.5.29} and configured to connect to a MySQL server. 
We collect end-user traffic performed through a web browser: 1) visit pages, 2) log in with a username and a password, 3) add some comments on the main page and on a specific user page, 4) update personal page information and 5) log out.
We record every HTTP request, as well as the associated HTTP responses (including response code, header and body).

In order to evaluate all the try-catch blocks in XWiki, this end-user traffic is replayed to perform each experiment as done in previous work~\cite{Breaking_a_production_brokerage_platform_without_causing_financial_devastation}. The reply is done on the production system directly, meaning that during the experiment, the system is still able to serve other requests as usual. This setup is also used for the evaluation of Broadleaf in \autoref{sec:evaluation-broadleaf}.

First, \chaosmachine runs the observation mode to monitor all the dynamic try-catch information and normal behavior without any perturbation. Then, an exploration mode is activated.
\chaosmachine activates the corresponding perturbation injector for each covered try-catch block. The injector is active for $1$ minute and \chaosmachine collects the HTTP responses, which are then compared to those collected in observation mode.

In XWiki's experiment, we define the four classes of try-catch blocks as follows:

\begin{itemize}
\item \emph{Resilient try-catch block}. Despite injected exceptions in this block, users still get the expected response content or succeed in adding comments and updating personal profile.
\item \emph{Observable try-catch block}. A try-catch block is said observable if the response code changes from ``200 OK'' to others. Consequently, users also get an error page or request redirection under the corresponding exceptions.
\item \emph{Debuggable try-catch block}. A try-catch block is said debuggable if the exception information is captured in application logs when an exception is injected.
\item \emph{Silent try-catch block}. A silent try-catch block only causes response body change while the response code stays the same as usual, and there is no error information about the injected exception in application logs.
\end{itemize}

\subsubsection{Experimental results}\label{sec:results-xwiki}

\begin{table}[!tbp]
\centering
\caption{Results on Chaos Experimentation on 268 Try-catch Blocks in XWiki Covered by the Considered Workload}\label{tab:tcInfoInXWiki}
\scriptsize
\begin{tabular}{lr p{1.8cm} rrrr}
\toprule
Packages& Covered& Executions in\newline Obse. / Expl.& RH& OH& DH& SH\\
\midrule
org/xwiki/a*& 1& 273 / 273& 0& 0& 1& 0\\
org/xwiki/c*& 20& 112968 / 119544& 0& 6& 20& 0\\
org/xwiki/d*& 2& 855 / 1398& 0& 0& 2& 0\\
org/xwiki/e*& 11& 20882 / 99204& 0& 1& 11& 0\\
org/xwiki/f*& 23& 44813 / 222& 0& 0& 23& 0\\
org/xwiki/i*& 8& 1142 / 280& 0& 0& 8& 0\\
org/xwiki/l*& 12& 295530 / 73048& 0& 1& 12& 0\\
org/xwiki/m*& 9& 38360 / 37739& 0& 1& 9& 0\\
org/xwiki/n*& 10& 62 / 190837& 0& 0& 8& 2\\
org/xwiki/o*& 2& 43753 / 68154& 0& 0& 2& 0\\
org/xwiki/p*& 4& 5403 / 3075& 0& 0& 4& 0\\
org/xwiki/q*& 3& 262 / 142& 0& 0& 3& 0\\
org/xwiki/r*& 93& 1137420 / \newline 272944& 5& 7& 70& 14\\
org/xwiki/s*& 15& 20522 / 31826& 2& 5& 15& 0\\
org/xwiki/t*& 2& 83 / 81& 0& 0& 2& 0\\
org/xwiki/u*& 20& 13795 / 6229& 0& 8& 16& 1\\
org/xwiki/v*& 5& 3201 / 831& 0& 2& 5& 0\\
org/xwiki/w*& 21& 2526 / 3140& 0& 2& 16& 5\\
org/xwiki/x*& 7& 890 / 580& 0& 0& 6& 1\\
\hline
Total& 268/1567& 1742740 / \newline 909547& 7& 33& 233& 23\\
\bottomrule
\end{tabular}
\end{table}

There are 290 user requests we recorded: 276 GET requests and 14 POST ones. This traffic contains: 97 GET requests directly on downloading resources, 178 GET requests and 10 POST requests on rendering pages, 4 POST requests on logging in, adding comments, updating user data, and 1 GET request on logging out.

In total, 1567 try-catch blocks are registered in \chaosmachine, and 268 of them are covered by the traffic we recorded.
\autoref{tab:tcInfoInXWiki} summarizes the data aggregated over packages.
The first column is the abbreviated package name.
The second column shows the number of try-catch blocks that are covered by the production traffic. The third column is the total number of try-catch block executions in both observation mode and exploration mode. Finally, the last four columns indicate the number of try-catch blocks which meet our pre-defined four hypotheses described in \autoref{sec:overview}, including:
7 resilient try-catch blocks,
33 observable try-catch blocks,
233 debuggable try-catch blocks,
and 23 silent try-catch blocks.

Take the row ``org/xwiki/s*'' as an example. For all the try-catch blocks in the package whose name begins with \texttt{org/xwiki/s}, there are 15 try-catch blocks covered by this set of chaos experiments. Under normal conditions, these 15 try-catch blocks are executed $20522$ times. When \chaosmachine activates the corresponding perturbation injectors in these try-catch blocks, the same blocks are executed $31826$ times in total. After classification by \chaosmachine, the developer knows that: 1) $2$ try-catch blocks satisfy the resilience hypothesis, 2) 5 try-catch blocks satisfy the observable hypothesis, 3) $15$ try-catch blocks satisfy the debug hypothesis and 4) none of the try-catch blocks satisfy the silence hypothesis.

With the help of this report, developers gain more knowledge on XWiki's error-handling capabilities in production. They are also encouraged to take action:
1) go over the silent try-catch blocks to confirm whether they need to record more information when an exception occurs and 2) focus on the try-catch blocks which have a serious impact on system's steady state, i.e. the observable ones. For example, if there is an exception in a specific try block, which leads to the system to generate an $500$ response code instead of $200$. As a result, the response contents also change to an error page for users. The chaos experiment provides more clues for developers to review the try-catch block and help them improve the fault tolerance ability.

\subsubsection{Case studies}

In the following, we detail two interesting cases found in the XWiki experiment.

\begin{lstlisting}[float,caption=XWikiException in XWikiCachingRightService/authenticateUser,label=lst:xwikiCase1]
try {
    XWikiUser user = context.getWiki().checkAuth(context);
    if (user != null) {
        userReference = resolveUserName(user.getUser(), new WikiReference(context.getWikiId()));
    }
} catch (XWikiException e) {
    LOGGER.error("Caught exception while authenticating user.", e);
}
\end{lstlisting}

\autoref{lst:xwikiCase1} shows part of method \texttt{authenticateUser} in class \texttt{XWikiCachingRightService}. There is only one try-catch block in this method. It is executed 151 times in observation mode and 153 times with perturbation. When the exception occurs, this catch block logs the error information. According to the monitored behavior, this perturbation actually has a visible impact on certain requests: it leads to an HTTP response code $302$ (Redirect) instead of $200$. Per our definition, this try-catch block satisfies both the observability and the debug hypothesis.

\begin{lstlisting}[float,caption=InterruptedException in DefaultSolrIndexer \$Resolver/runInt- ernal,label=lst:xwikiCase2]
try {
    queueEntry = resolveQueue.take();
} catch (InterruptedException e) {
    logger.warn("The SOLR resolve thread has been interrupted", e);
    queueEntry = RESOLVE_QUEUE_ENTRY_STOP;
}

if (queueEntry == RESOLVE_QUEUE_ENTRY_STOP) {
    // Stop the index thread: clear the queue and send the stop signal without blocking.
    indexQueue.clear();
    indexQueue.offer(INDEX_QUEUE_ENTRY_STOP);
    break;
}
\end{lstlisting}

\autoref{lst:xwikiCase2} shows part of method \texttt{runInternal} in class \texttt{DefaultSolrIndexer}'s private inner class \texttt{Resolver}. This try block is executed 11 times in observation mode and is executed only once with perturbation. \chaosmachine identifies that this perturbation does not influence the output of any request. The monitoring sidecar also detects that the exception is caught in the application log. As we can see from the source code, developers log the exception information and also assign \texttt{queueEntry} to \texttt{RESOLVE\_QUEUE\_ENTRY\_STOP} in the catch block which is a valid error-handling strategy in this context. Through the chaos experiment, the developers gain more confidence that this exception-handling design actually works in production.

\begin{mdframed}
Main result of the XWiki experiment: \chaosmachine analyzes 268 try-catch blocks and identifies 7 that satisfy the resilience hypothesis, and 23 try-catch blocks that are silent, violating the silence hypothesis. This experiment shows that our prototype implementation of \chaosmachine scales to a system with $440k$ LOC and 1567 try-catch blocks loaded in the JVM.
\end{mdframed}

\subsection{Evaluation on Broadleaf}\label{sec:evaluation-broadleaf}

\subsubsection{Introduction of Broadleaf}
\href{https://www.broadleafcommerce.com/}{Broadleaf Commerce} is a series of open-source products in an e-commerce platform written in Java. There are three components in Broadleaf which can be deployed separately into different servers: administration website, end-user shopping website and data fetching APIs.

\subsubsection{Experiment protocol}\label{experimentProtocolBroadleaf}
We choose to conduct chaos experiments on Broadleaf version \href{https://github.com/BroadleafCommerce/LegacyDemoSite/tree/broadleaf-5.0.0-GA}{5.0.0-GA}. It provides an embedded Tomcat server, a HyperSQL database and a startup script, which simplifies deployment. For this experiment, we focus on the end-user shopping website. Similar to the experiments on XWiki, we deploy Broadleaf and randomly interact with the website system, including: 1) visiting product pages, 2) logging in with a username and a password, 3) adding products to a shopping cart, 4) checking out, and 5) logging out. As before, we record every user request and its associated response.
In this experiment, we define resilient, observable, debuggable, silent try-catch block as per the XWiki experiment in \autoref{sec:experimentProtocolXWiki}, since they are both web systems with the same core characteristics.

\subsubsection{Experimental results}\label{sec:results-broadleaf}

The recorded operations include $384$ requests in total. There are $362$ requests responsible for directly downloading files, all of which are GET requests. There are $15$ GET requests about rendering pages. All of the 6 functional requests are POST, including logging in, updating the shopping cart, and checking out. The request for logging out is a request of type GET. These requests are replayed by GoReplay all the time during the experiments, and the time to finish this sequence of operations is less than $90$ seconds. First, \chaosmachine keeps a $90$ seconds observation mode to gather the system's normal behaviors. At the same time, it also obtains information about covered try-catch blocks. Then, for each covered try-catch block, \chaosmachine runs in exploration mode for $90$ seconds. The results are generated and discussed next.

\begin{table}[!tbp]
\caption{Results on Chaos Experimentation on 44 Try-catch Blocks in Broadleaf Covered by the Considered Workload}\label{tab:tcInfoInBroadleaf}
\centering
\scriptsize
\begin{tabular}{p{2.2cm}r p{1.5cm} rrrr}
\toprule
Packages& Cove.& Executions in\newline Obse. / Expl.& RH& OH& DH& SH\\
\midrule
o/b/cms/file*& 1& 53 / 50& 0& 1& 1& 0\\
o/b/cms/url*& 3& 288 / 111& 2& 0& 0& 1\\
o/b/com*/audit*& 2& 40 / 13& 0& 0& 1& 1\\
o/b/com*/classloader*& 2& 1596 / 849& 0& 2& 2& 0\\
o/b/com*/i18n*& 1& 10660 / 51& 0& 1& 1& 0\\
o/b/com*/persistence*& 1& 24 / 2& 0& 0& 1& 0\\
o/b/com*/security*& 2& 14 / 40& 0& 1& 2& 0\\
o/b/com*/util*& 1& 30 / 21& 0& 1& 1& 0\\
o/b/com*/web*& 4& 188 / 60& 0& 2& 3& 1\\
o/b/core/catalog*& 1& 2 / 2& 0& 0& 0& 1\\
o/b/core/order*& 5& 34 / 84& 0& 3& 5& 0\\
o/b/core/payment*& 1& 1 / 1& 1& 0& 0& 0\\
o/b/core/pricing*& 1& 5 / 21& 0& 1& 1& 0\\
o/b/core/rating*& 2& 6 / 6& 0& 0& 2& 0\\
o/b/core/search*& 2& 44 / 38& 0& 2& 2& 0\\
o/b/core/web*& 10& 615 / 340& 1& 5& 7& 2\\
o/b/ope*/audit*& 3& 16 / 14& 0& 1& 1& 2\\
o/b/profile/core*& 1& 3 / 2& 1& 0& 0& 0\\
o/b/vendor/sample*& 1& 1 / 1& 0& 1& 1& 0\\
\hline
Total& 44/355& 13620 / 1706& 5& 21& 31& 8\\
\bottomrule
\end{tabular}
\end{table}

\autoref{tab:tcInfoInBroadleaf} summarizes the results. The recorded traffic covers $44$ try-catch blocks. In the first step of the experiment, we leave \chaosmachine running automatically. In this case, it does not detect any resilient try-catch blocks, which leads us to do some further analysis. In the second step, we manually analyze all logs generated by monitoring sidecars. This analysis reveals that some of the diff-logs are semantically equivalent, but the monitoring sidecar marks the output as different if it is not the same.

For instance, Broadleaf uses JSON objects to handle the prices of products with different properties. The price of an XL-size black T-shirt is \$17, which is displayed as \texttt{\{"options":[1, 14], "price":17\}}. In the snippet, number 1 stands for ``XL'' and number 14 stands for ``black''. It is obvious that \texttt{\{"options":[14, 1], "price":17\}} has the same meaning. However, the current implementation of our monitoring sidecar regards these as different outputs. This phenomenon reflects one limitation of the monitoring sidecar: it is not sophisticated enough to determine semantical equivalence.

Following the manual comparison between the response bodies, the revised report about try-catch resilience is:
5 resilient try-catch blocks,
21 observable try-catch blocks,
31 debuggable try-catch blocks,
and 8 silent try-catch blocks.

\subsubsection{Case studies}
Next, we discuss one of the most interesting cases found in the chaos experiment on Broadleaf.

\begin{lstlisting}[float,caption=NoResultException in CountrySubdivisionDaoImpl/findSubdivisionByCountryAndAltAbbreviation,label=lst:blcCase1]
public CountrySubdivision findSubdivisionByCountryAndAltAbbreviation(...) {
    TypedQuery<CountrySubdivision> query = new ...
    try {
        return query.getSingleResult();
    } catch (NoResultException e) {
        return null;
    }
}
\end{lstlisting}

As shown in~\autoref{lst:blcCase1}, \chaosmachine identifies this try-catch block as a resilient one. As the method name suggests, the method is used for obtaining the sub-division of a country. The method is executed $3$ times in observation mode and $2$ times in exploration mode. When the perturbation injector is activated, the query result is always ``null''. The reason why the response content stays the same is that in observation mode, the user's country information does not contain sub-divisions. Thus, no matter if there is an exception, the query result remains ``null''. However, this try-catch block may impact the system's output if a specific user has sub-division information.

This phenomenon exposes another limitation of \chaosmachine. Since it uses production traffic to evaluate the resilience of try-catch blocks, the traffic might not be sufficient to give definitive conclusions. For some try-catch blocks that are currently classified as resilient, different traffic may be able to falsify the hypotheses. The accuracy of the report of \chaosmachine can be optimized by using more varied production traffic.

\subsubsection{Discussion of Broadleaf experiment}
The experiments on Broadleaf reveal two limitations of \chaosmachine as discussed above: 1) the monitoring sidecar has no automated way of semantically comparing the outputs (in this case the JSON outputs), hence, in addition to the automated capabilities of \chaosmachine further manual work may be needed to improve monitoring; 2) when the availability of production traffic is limited, \chaosmachine only guarantees that the try-catch identifications are correct under the considered workload. We note that it is also possible to generate more diverse traffic or capture the production traffic for a longer time.

\begin{mdframed}
Main result of the Broadleaf experiment: \chaosmachine identifies $5$ resilient, $8$ silent ones, $21$ observable ones, and $31$ try-catch blocks. This experiment exposes two important facts: (a) the monitoring sidecar may need to embed some domain-specific user-defined metrics in order to better interpret the application output and logs, and (b) the length of the captured production traffic during chaos experimentation matters.
\end{mdframed}

\subsection{Overhead of the \chaosmachinebf}
Now, we discuss the overhead of \chaosmachine. 
We present in the context of TTorrent because the overall downloading time of a file is a clear cut metric. The results for the other case studies are equivalent. We calculate the overhead of \chaosmachine in three aspects: 1) at the application level, by measuring downloading time increase between the original version and the instrumented version, 2) at the system level, by measuring CPU and memory usage increase, and 3) at the binary code level, by measuring code bloat due to instrumentation. For statistical purposes, we conduct the same measurement 5 times and calculate the average. The results are presented in~\autoref{tab:overheadOfTTorrent}.

\begin{table}[!tbp]
\centering
\caption{The Overhead of An Exploration Experiment on TTorrent}\label{tab:overheadOfTTorrent}
\scriptsize
\begin{tabular}{lrrr}
\toprule
Evaluation Aspects& Original Version& Instrumented Version& Variation\\
\midrule
Downloading time& 102.2s& 96.4s& -6\%\\
CPU time& 2288& 3410& 50\%\\
Memory usage& 289M& 332M& 15\%\\
Peak thread count& 119& 116& -3\%\\
Class files size& 330.Kb& 334.Kb& 1.3\%\\
\bottomrule
\end{tabular}
\end{table}

The instrumentation done by \chaosmachine has little influence on downloading time, memory usage and file size. We observe a 50\% higher CPU time; this is due to the fact that \chaosmachine turns on all the monitoring sidecars to print more information. Note that in falsification mode, when \chaosmachine focuses on some specific hypotheses, the overhead of CPU time is significantly reduced to less than 1\%. As a summary, the overhead of chaos experiments on TTorrent can be considered as compatible with production requirements.

\section{Discussion}\label{sec:discussion}
\subsection{Threats to Validity}
\subsubsection{Internal Validity}
The main threat to internal validity is that changing the perturbation model may change the results of the experiment. For example, another perturbation model could inject exceptions at the end of a try-catch block instead of at the beginning. Future work will analyze how different perturbation models influence the experimental results.

\subsubsection{External Validity}
\chaosmachine has been evaluated on three different kinds of Java applications, including one client-side file downloading tool and two server-side web applications. The validity for other application domains and languages which support the JVM as compilation target has to be verified. Further work will evaluate the application of \chaosmachine to JVM bytecode compiled from other languages like Scala and Kotlin. It would also be interesting to study \chaosmachine in the context of application frameworks like Spring\cite{Johnson:2005:Professional_Java_Development_with_the_Spring_Framework} and Akka\footnote{\url{https://akka.io/}} for reactive applications.

\subsubsection{Construct Validity}
One of the threats to construct validity is that \chaosmachine does not consider the potentially delayed causality of an injected exception. As described in \autoref{sec:evaluation}, \chaosmachine only turns on one perturbation injector during an experiment. If the perturbation has a direct influence on the application's behavior, \chaosmachine detects the side-effects for the current request. However, \chaosmachine is not able to analyze whether a triggered error has an impact on subsequent requests. In future work, this limitation could be addressed 1) by defining specific hypotheses regarding an application's behavior over time, and 2) by changing the granularity of a single chaos experiment.

\subsection{Extensibility of \chaosmachinebf}
\chaosmachine is designed to evaluate the error-handling capabilities of Java applications. More specifically, it focuses on the resilience of try-catch blocks. Yet, it is possible to customize \chaosmachine for other fault injection experiments. Indeed, \chaosmachine provides an interface which defines a byte code instrumentation method \texttt{generateByteCode}. By default, \chaosmachine initializes a default perturbation injector as explained in~\autoref{sec:perturbationInjectors}.

This makes \chaosmachine extensible: developers are free to implement their own perturbation strategy by implementing this interface. To prevent unwanted interactions between different perturbation strategies, \chaosmachine instantiates a single strategy per experiment. This does not impact the monitoring sidecar, the metrics are systematically extracted from Java Management Extensions (JMX) for all new perturbators. Yet, developers can also define new application-level metrics. In summary, we have designed \chaosmachine with the intention to reduce the effort for customization.

\section{Related work}\label{sec:relatedWork}
Chaos engineering {\em per se} is a new field which is little researched, hence the closely related work is relatively scarce. Beyond chaos engineering, we discuss complementary work in the areas of fault-injection, static analysis of error-handling, and dependability benchmarking.

\subsection{Software Fault Injection}
Software fault injection (SFI) is a well-researched area in the field of software dependability; we refer to Natella et al.'s ~\cite{natella2016assessing} survey about SFI including its concepts, applications and comparisons. It is traditionally applied offline to evaluate error-handling capabilities. Kanawati et al.~\cite{Ghani:FERRARI} proposed FERRARI, a tool for the validation of dependability properties. Han et al.~\cite{Seungjae:DOCTOR} designed DOCTOR, an integrated environment for assessing distributed real-time systems, and Lee et al.~\cite{Hyosoon:SFIDA_a_software_implemented_fault_injection_tool_for_distributed_dependable_applications} proposed SFIDA, a tool to test the dependability of distributed applications on the Linux platform. All of these tools are based on injecting hardware-related faults, in a testing setup.
Montrucchio et al.~\cite{Montrucchio:Fault_injection_in_the_process_descriptor_of_a_Unix-based_operating_system}, Segal et al.~\cite{Barton:Fault_injection_experiments_using_FIAT,Segall:FIAT-fault_injection_based_automated_testing_environment}, Arlat et al.~\cite{Jean:Fault_Injection_for_Dependability_Validation} also presented similar injection techniques for simulating hardware faults. Kao et al.\cite{Kao:A_fault_injection_and_monitoring_environment_for_tracing_the_UNIX_system_behavior_under_faults} invented ``FINE'', a fault injection and monitoring tool to inject both hardware-induced software errors and software faults. Cotroneo et al. proposed a methodology to assess OpenStack's resilience with respect to software failures using fault injection~\cite{Cotroneo:An_Empirical_Analysis_of_Software_Failures_in_the_OpenStack_Cloud_Computing_Platform}. Ghidei~\cite{Ghidei:LDFI_for_actor_based_programs} proposed LDFI-Akka, a tool that employs lineage-driven fault injection for actor-based programs to analyze the weaknesses of Akka programs.
All those systems are not meant to be used in production, because, in the literature, fault tolerance analysis is done at design or testing time.
More importantly, it is actually not possible to use them in production out-of-the-box, either because they require the source code or because they impose an unacceptable overhead. In contrast, \chaosmachine is designed for fault injection in production in order to give precious insights of error-handling capabilities in a live setting.

Marinescu and Candea described LFI~\cite{Marinescu:Efficient_Testing_of_Recovery_Code_Using_Fault_Injection}, a reusable and scalable library-level fault injection framework to test the recovery code of a given system. The common idea of LFI and \chaosmachine is that neither of them requires the source code, and that fault injection happens at runtime. However, there are two main differences between these techniques: 1) LFI injects failures into common libraries on which an application depends, while \chaosmachine injects exceptions directly into the application; 2) LFI injects faults by manipulating error return codes and corresponding side effects while \chaosmachine generates application-level exceptions.

Netflix~\cite{Chang:Chaos_Monkey_Increasing_SDN_Reliability_Through_Systematic_Network_Destruction} is well known for its ChaosMonkey, which randomly shuts down Amazon instances in production. It is used to ensure that the user experience is not impacted by a loss of an Amazon instance. This methodology has been extended to more failure types both at Netflix~\cite{The_Netflix_Simian_Army} and other companies~\cite{Heather:Inside_Azure_Search_Chaos_Engineering}. An example of cloud-oriented tool is by Sheridan et al.~\cite{Craig:DICE_Fault_Injection_Tool}, who presented a fault injection tool for cloud applications, where faults are resource stress or service outage.
While those tools conduct chaos experiments between services at the OS level or the network level, \chaosmachine is, to the best of our knowledge, the first to perform chaos experiments in a white-box fashion. It perturbs the runtime status inside the JVM, and enables developers to detect internal weaknesses at the code level (not the service interaction level).

\subsection{Exception Analysis}
Now we discuss exception analysis.
Martins et al.~\cite{Alexandre:Testing_Java_Exceptions} presented VerifyEx to test Java exceptions by inserting exceptions at the beginning of try blocks. Their goal is to improve test coverage and not to assess error-handling contracts as done in \chaosmachine.
Byeong-Mo et al.~\cite{Byeong-Mo:A_review_on_exception_analysis} gave a comprehensive review on exception analysis.
Fu and Ryder\cite{Fu:Exception-Chain_Analysis} described a static analysis method for exception chains in Java.
Magiel Bruntink et al.~\cite{Magiel:Discovering_faults_in_idiom-based_exception_handling} proposed a characterization and evaluation method to discover faults in idiom-based exception handling.
Zhang and Elbaum\cite{Zhang:Amplifying_Tests_to_Validate_Exception_Handling_Code} presented an approach that amplifies test to validate exception handling. Cornu et al.\cite{Cornu:Exception_Handling_Analysis_and_Transformation_Using_Fault_Injection} proposed a classification of try-catch blocks at testing time.
Here, the problem domain and implementation techniques are different: those authors use modified source code and test suites to study resilience. On the contrary, \chaosmachine operates in production with Java bytecode, using real production traffic to conduct the analysis.

Czeck et al.\cite{Czeck:Observations_on_the_effects_of_fault_manifestation_as_a_function_of_workload} described a methodology for modeling fault effects on system behavior. They construct a behavior model based on a small set of workloads and use the model to infer the fault behavior of other workloads. In comparison, \chaosmachine is directly applied to the production system to make and falsify hypotheses about its resilience.
Finally, chaos engineering relates to failure-oblivious computing \cite{rinard2004enhancing}: both are engineering techniques for production failures, yet failure-oblivious computing is not about the active injection of faults in the production systems.

\subsection{Dependability Benchmarking}
Dependability benchmarking is another relevant field. This is a systematical process to characterize system behavior in the presence of faults ~\cite[p. xiii]{Kanoun:Dependability_Benchmarking_for_Computer_Systems}. Kanoun et al. \cite{kanoun:DBench} proposed a framework called DBench, which defines a series of benchmarks for off-the-shelf software components. Dur{\~{a}}es et al. \cite{Joao:Dependability_Benchmarking_of_Web-Servers} designed a dependability benchmark for web-servers. Sangroya and Bouchenak \cite{Amit:A_Reusable_Architecture_for_Dependability_and_Performance_Benchmarking_of_Cloud_Services} presented a generic software architecture for dependability and performance benchmarking of cloud computing services. Sangroya et al. \cite{Amit:Experience_with_benchmarking_dependability_and_performance_of_MapReduce_systems} proposed specifically a benchmark suite for evaluating the dependability and performance of MapReduce~\cite{DeanG08} systems. Herscheid et al.~\cite{Lena:Hovac_A_Configurable_Fault_Injection_Framework_for_Benchmarking_the_Dependability_of_C_C++_Applications} invented Hovac, a configurable tool for dependability benchmarking of C/C++ applications.

Both \chaosmachine and the above dependability benchmarking frameworks use fault injection techniques to perturb a software system. However, the methodologies are very different. The main difference is that \chaosmachine takes advantage of production workloads to evaluate the given application. Moreover, \chaosmachine defines application-specific hypotheses before experiments, while benchmarks usually conduct generic measurements based on implicit oracles (in particular crashes).

\section{Conclusion}\label{sec:conclusion}
This paper presented \chaosmachine, which analyzes and falsifies exception-handling hypotheses in Java programs running in production. We showed, on three large applications, that \chaosmachine is able to produce actionable reports for developers to gain more confidence about the resilience of their system, and to point out critical try-catch blocks that need more attention.
In future work, we will improve the monitoring sidecar to capture more precisely the steady state of the system. We will also design more advanced perturbation models, for example by changing the timing of methods invocation.

\section*{Acknowledgements}
This work was partially supported by the Wallenberg AI, Autonomous Systems and Software Program (WASP) funded by the Knut and Alice Wallenberg Foundation.

\bibliographystyle{plain}
\bibliography{references}

\begin{IEEEbiography}[{\includegraphics[width=1in,height=1.25in,clip,keepaspectratio]{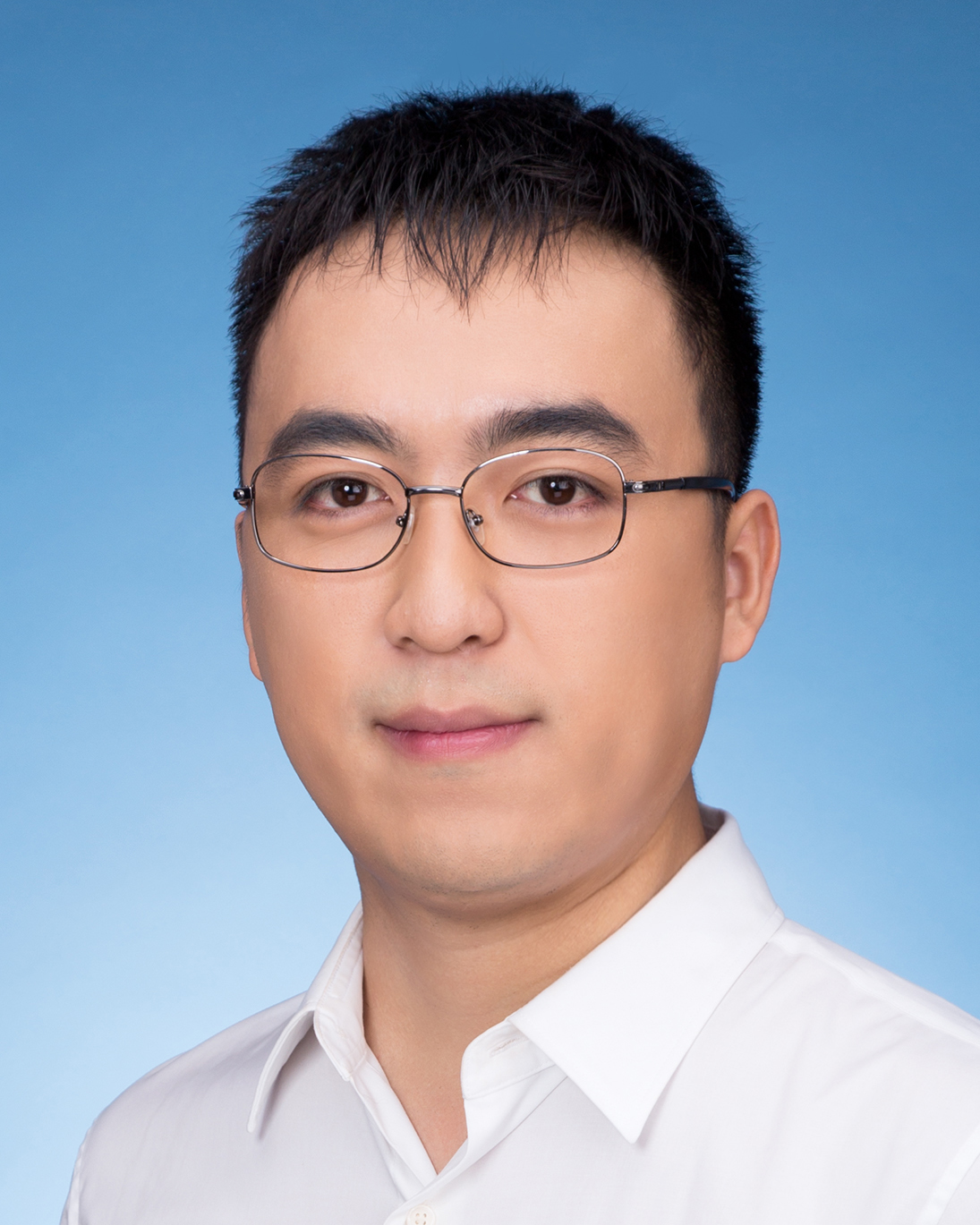}}]{Long Zhang}
is now a Ph.D. student in computer science at KTH Royal Institute of Technology, Sweden. His research work focuses on self-healing software, chaos engineering and antifragile systems. Long received his BE degree and ME degree in software engineering from Harbin Institute of Technology, China. Before his Ph.D. study, Long was hired by Tencent as a software developer and project manager, who was responsible for university-enterprise cooperation projects design and development.
\end{IEEEbiography}
\vskip -1\baselineskip

\begin{IEEEbiography}[{\includegraphics[width=1in,height=1.25in]{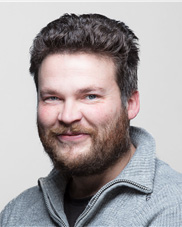}}]{Brice Morin}
is a Senior Research Scientist at SINTEF Digital, Oslo, Norway. His research focuses on Model-Driven Software Engineering for complex, heterogeneous and distributed systems. He is a core contributor to ThingML, an open-source high-level programming language targeting this kind of systems. He received a Ph.D. from the University of Rennes (France) in 2010.
\end{IEEEbiography}
\vskip -1\baselineskip

\begin{IEEEbiography}[{\includegraphics[width=1in,height=1.25in]{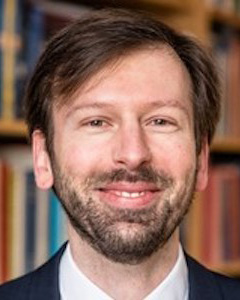}}]{Philipp Haller}
is an Associate Professor of Computer Science at KTH Royal Institute of Technology (Sweden). He was part of the team that received the 2019 ACM SIGPLAN Programming Languages Software Award for the development of the Scala programming language. He received a Ph.D. from EPFL (Switzerland) and a Diplom-Informatiker degree from Karlsruhe Institute of Technology (Germany). His main research interests are in the design and implementation of programming languages, type systems, concurrency, and distributed programming.
\end{IEEEbiography}
\vskip -1\baselineskip

\begin{IEEEbiography}[{\includegraphics[width=1in,height=1.25in]{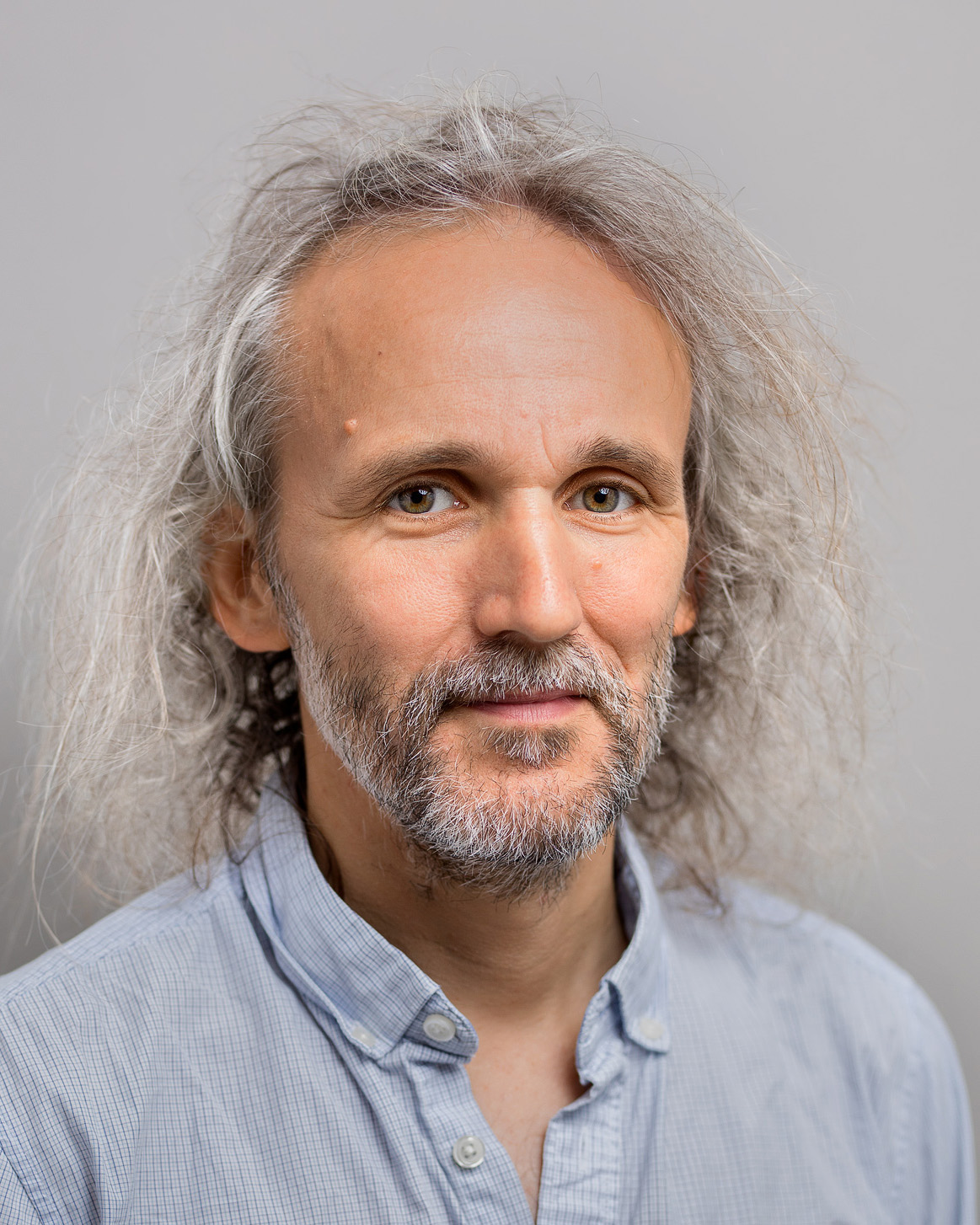}}]{Benoit Baudry}
is a WASP Professor at the KTH Royal Institute of Technology in Stockholm, Sweden, and the director of the CASTOR center for software research at KTH. Until August 2017 he was a research scientist at INRIA in Rennes, France, where he led the DiverSE research group (EPI) from 2013 to 2017. His research focuses on software diversification and testing for reliability and moving target defenses. He performs experimental research with large open source software systems. Experiments support sound science and close collaboration with software industry.
\end{IEEEbiography}
\vskip -1\baselineskip

\begin{IEEEbiography}[{\includegraphics[width=1in,height=1.25in]{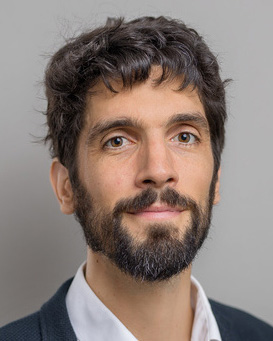}}]{Martin Monperrus}
is Professor of Software Technology at KTH Royal Institute of Technology. He was previously associate professor at the University of Lille and adjunct researcher at Inria. He received a Ph.D. from the University of Rennes, and a Master's degree from the Compiègne University of Technology. His research lies in the field of software engineering with a current focus on automatic program repair, program hardening and chaos engineering. 
 \end{IEEEbiography}
\vfill

\end{document}